\pdfoutput=1
\documentclass[aps,prl,reprint,groupedaddress,floatfix]{revtex4-1}

\usepackage{graphicx}% Include figure files
\usepackage{dcolumn}% Align table columns on decimal point
\usepackage{bm}% bold math
\usepackage{physics}
\usepackage{amssymb}
\usepackage{amsmath}
\begin{document}

%\preprint{}

\title{Shape Recovery in Viscoelastic Silicone Rubber and the Fractional Zener Model}


\author{Louis A. Bloomfield}
\email{lab3e@virginia.edu}
%\thanks{}
\affiliation{Department of Physics, University of Virginia, Charlottesville, VA 22904}


\date{\today}

\begin{abstract}
Viscoelastic silicone rubber (VSR) is a remarkable shape-memory solid. The material's polymer network retains a memory of its shape history, so its current and future shapes depend strikingly on its past shapes. Although VSR's memory fades gradually and it has a permanent (cured-in) shape to which it will eventually return when left alone, VSR can be taught new shapes and retain them for significant lengths of time.

To examine VSR's ability to learn, remember, and recover shapes, this work focuses on a simple experiment. A VSR that has relaxed into its permanent shape is suddenly compressed to about 80\% of its original height. After a specific period of compression, the VSR is released and allowed to return to its permanent shape. Having learned a new shape during the compression period, however, the VSR is reluctant to return and takes seconds, minutes, or hours to do so, depending on how long it was compressed.

In addition to observing these behaviors experimentally in VSR, we show that those behaviors are well-described by a simple viscoelastic model. Unlike typical viscoelastic models, which are constructed from integer-order viscoelastic elements (e.g. elastic springs and viscous dashpots), the model describing VSR is the Fractional Zener model and involves a fractional-order element known as a spring-pot. Here ``fractional'' refers to the branch of mathematical analysis known as fractional calculus, a discipline that deals with derivatives, integrals, and differential equations of non-integer order. For example, between the first derivative and a second derivative, there are an infinite number of fractional derivatives. Though well-developed and important, fractional calculus is far less familiar than integer calculus, so this article is necessarily somewhat pedagogical.

For integer-order viscoelastic models and the materials they describe, the future depends only on the present. For fractional-order models, the future depends also on the past. That memory of the past is intrinsic to fractional time derivatives: the fractional time derivative of any function $f(t)$ depends not only on $f(t')$ at times $t'$ infinitesimally close to time $t$, but also on $f(t')$ at all times $t'$ where $t' < t$.

Both VSR and the Fractional Zener model that describes its behaviors are acutely aware of the past. The model's mathematical machinery make it possible to design VSR behaviors based on physical parameters, although some of the model's relationships are not yet known in closed form. VSR's existence as a practical material means that devices can be designed and produced that use a memory of past shapes to do things that would otherwise be difficult or impossible to make.
\end{abstract}

%\keywords{}

\maketitle

\section{Introduction}
Viscoelastic silicone rubber (VSR) is a unique shape-memory solid. Its shape-memory allows VSR to temporarily adopt new shapes imposed on it by its environment but gradually recovers its permanent equilibrium shape when freed of external forces. That it learns new shapes makes VSR well-suited to a broad range of padding and supporting applications. That it returns to its permanent shape when freed from constraints makes it great for many sealing applications.

VSR's remarkable elastic and viscoelastic behaviors derive from in its unusual polymer network. Its silicone polymer chains are joined together by both permanent \textit{and} temporary crosslinks. While all of its crosslinks involve strong covalent bonds, the temporary crosslinks detach and reattach frequently and thus have finite lifetimes. Because VSR's permanent crosslink concentration exceeds the gelation threshold, VSR exhibits the elastic characteristics of a network solid. Because its temporary crosslink concentration gives rise to marked time dynamics, VSR also exhibits the viscoelastic characteristic of a network liquid. 

\begin{figure}
	\includegraphics[width=0.4\columnwidth]{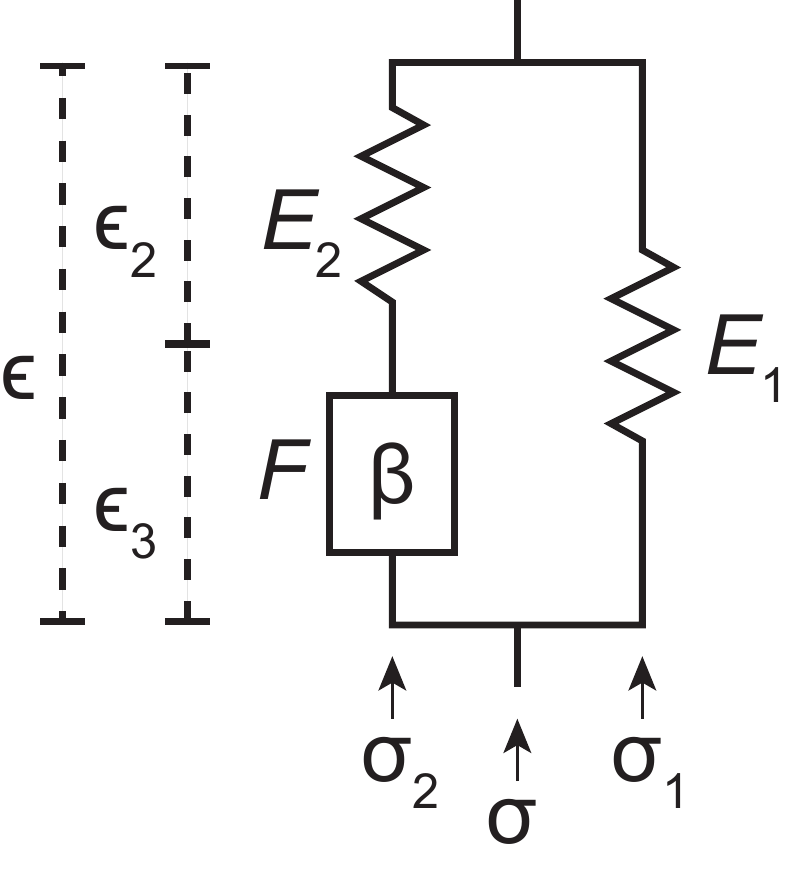}%
	\caption{The Fractional Zener viscoelastic model consists of two springs, $E_{1}$ and $E_{2}$, and fractional-order viscoelastic element known as a spring-pot, $F$. The stresses $\sigma$ and strains $\epsilon$ shown here are used in Eqs. \ref{eq:SE1}--\ref{eq:SE5}.\label{fig:FractionalZener}}
\end{figure}

In previous work\cite{bloomfield2018} it was shown that when a VSR is subjected to sudden change in strain, its stress relaxation is well-described by the Fractional Zener model, a simple viscoelastic model of fractional order (Fig. \ref{fig:FractionalZener}). In this work it is shown that when a strained VSR is released from stress, its shape recovery is also described by the Fractional Zener model, but with a different characteristic time.

\section{The Fractional Zener Model}

The Fractional Zener viscoelastic model has three elements: two springs and a spring-pot. While springs are familiar viscoelastic elements, a spring-pot is a viscoelastic element of fractional order and, as such, needs introduction. 

Consider first a spring and a dashpot. These two are ordinary integer-order viscoelastic elements and their infinitesimal stress and strain are related by equations involving the ordinary differentiation operator
\begin{equation}
D^{n}_{x}\equiv \frac{d^{n}}{dx^{x}},
\end{equation}
where $n$ is an integer.
For a spring, 
\begin{equation}
\sigma(t) = E D^{0}_{t}\epsilon(t),\label{eq:ESpring}
\end{equation}
where $E$ is the spring's modulus. For a dashpot, 
\begin{equation}
\sigma(t) = \eta D^{1}_{t}\epsilon(t),\label{eq:EDashpot}
\end{equation}
where $\eta$ is the dashpot's viscosity.

Because a spring-pot is a fractional-order viscoelastic element, however, the equation relating its stress and strain involves a generalized differentiation operator
\begin{equation}
D^{\alpha}_{x}\equiv \frac{d^{\alpha}}{dx^{\alpha}},
\end{equation}
where $\alpha$ is not necessarily an integer. $D^{\alpha}_{x}$ belongs to the rich and well-developed branch of mathematical analysis known as fractional calculus.\cite{oldham1974,rabotnov1980,podlubny1998} Written in terms of that operator, the spring-pot's stress-strain equation is
\begin{equation}
\sigma(t)=F D^{\beta}_{t}\epsilon(t),\label{eq:ESpringpot}
\end{equation}
where $1 \ge \beta \ge 0$ is the fractional-order of the spring-pot and $F$ is its viscoelastic modulus. Note that a spring-pot becomes an ordinary dashpot when $\beta=1$ and an ordinary spring when $\beta=0$. 

While the operator $D^{\alpha}_{x}$ must reduce to $D^{n}_{x}$ whenever $\alpha$ is an integer $n$, $D^{\alpha}_{x}$ is not uniquely defined for non-integer values of $\alpha$. There are numerous definitions for $D^{\alpha}_{x}$ in the literature.\cite{deOliveira2014} The definition we will use here is the Riemann-Liouville left-sided derivative\cite{podlubny1998,deOliveira2014},
\begin{equation}
{}_{a}\textbf{D}_{x}^{\alpha}f(x) \equiv \frac{1}{\Gamma(n-\alpha)}\frac{d^{n}}{dx^{n}}\int_{a}^{x}(x-\xi)^{n-\alpha-1}f(\xi)d\xi\label{eq:RLLS}
\end{equation}
for  $x \ge a$ and $n$ is an integer such that $n-1 \le \alpha < n$. The Riemann-Liouville left-sided derivative operator's lower terminal $a$ defines the interval $[a,x]$ that the operator considers. In effect, that interval is the fractional derivative's ``memory'' and $a$ is usually chosen to be before anything of importance has occurred. It can even be chosen to be $-\infty$.

One peculiarity of ${}_{a}\textbf{D}^{\alpha}_{x}$ is that it yields a non-zero value when applied to a constant. There are other fractional derivative operators that eliminate that behavior,\cite{jumarie2007fractional} but we will find it sufficient to recognize when ${}_{a}\textbf{D}^{\alpha}_{x}$ is operating on a constant and set the result equal to zero.

With its two springs and one spring-pot, the Fractional Zener Model gives rise to five simultaneous equations
\begin{align}
\epsilon(t)=\epsilon_{2}(t)+\epsilon_{3}(t)\label{eq:SE1}\\
\sigma(t)=\sigma_{1}(t)+\sigma_{2}(t)\label{eq:SE2}\\
\sigma_{1}(t)=E_{1} \epsilon(t)\label{eq:SE3}\\
\sigma_{2}(t)=E_{2} \epsilon_{2}(t)\label{eq:SE4}\\
\sigma_{2}(t)=F {}_{a}\textbf{D}^{\beta}_{t}\epsilon_{3}(t)\label{eq:SE5}
\end{align}
where $t \ge a$, $1 \ge \beta \ge 0$, $\epsilon_{2}$ is the strain in spring $E_{2}$, $\epsilon_{3}$ is the strain in the spring-pot, $\sigma_{1}$ is the stress in spring $E_{1}$, and $\sigma_{2}$ is the stress in both spring $E_{2}$ and the spring-pot. In Section B below, those five equations are combined to eliminate the two internal stresses and two internal strains and obtain the fractional differential equation for the Fractional Zener model,
\begin{multline}
\sigma(t)+\frac{F}{E_{2}}{}_{a}\textbf{D}^{\beta}_{t}\sigma(t)\\
=E_{1}\epsilon(t) + F \left(\frac{E_{2}+E_{1}}{E_{2}}\right){}_{a}\textbf{D}^{\beta}_{t}\epsilon(t),\label{eq:FZMDE}
\end{multline}
where $t \ge a$.

\section{Stress Relaxation in VSR}

When VSR is subject to an instantaneous step in strain at time $t_{0}$, it responds with an instantaneous step in stress and then gradually relaxes monotonically toward a smaller constant-valued static stress. 
That relaxation process can be characterized by the VSR's stress relaxation modulus $G(t-t_{0})$.\cite{shaw2012,bloomfield2018}

To measure $G(t-t_{0})$, a 9.53mm-diameter cylinder of VSR is compressed from 6.35mm tall to 4.98mm tall in 10ms and held at that height while a load cell records the compressive force on the cylinder at 8ms time intervals. Dividing the compressive force by the cylinder's cross sectional area yields the compressive stress $\sigma_{c}(t)$. Since the compression is not infinitesimal, obtaining $G(t-t_{0})$ from $\sigma_{c}(t)$ requires the constitutive equation for a viscoelastic material\cite{bloomfield2018}
\begin{equation}
\sigma_{c}(t)=-\left(\lambda^{2}-\frac{1}{\lambda}\right)G(t-t_{0})\label{eq:StressConst}
\end{equation}
where $\lambda$ is the ratio of the cylinder's final height to its initial height and both the Finger tensor for compressive strain and the Poisson's ratio of rubbers (0.5) have been used.

\begin{figure}
	\includegraphics[width=0.8\columnwidth]{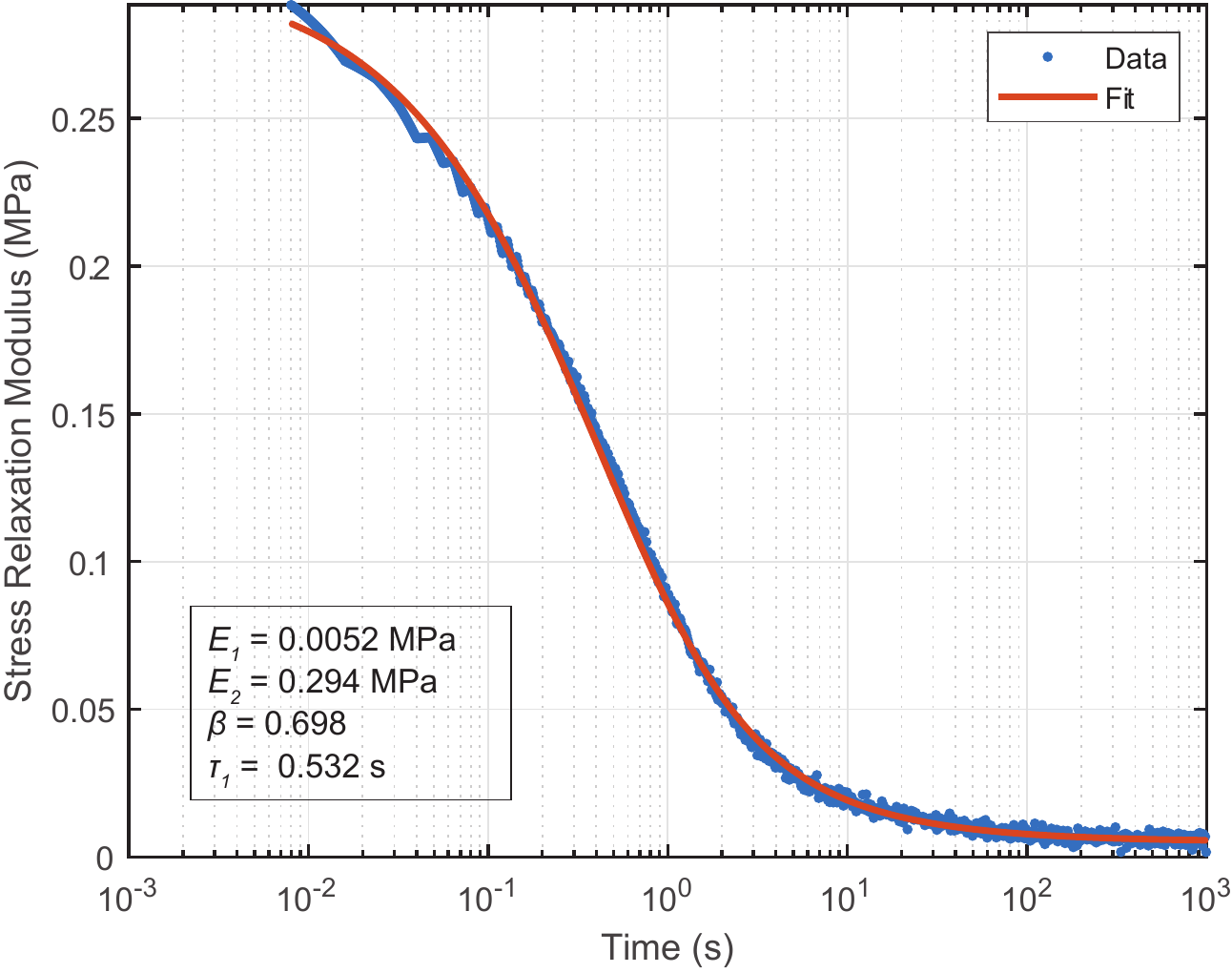}%
	\caption{Stress relaxation modulus $G(t-t_{0})$ of 190806AB VSR, measured via the sudden compression technique. The data are fit to the Fractional Zener model, as represented by Eq. \ref{eq:SRSol}. Static modulus $E_{1}$, transient modulus $E_{2}$, fractional order $\beta$, and characteristic time $\tau_{1}$ are the model and equations' four parameters.\label{fig:Compression}}
\end{figure}

Figure \ref{fig:Compression} shows $G(t-t_{0})$ obtained in this manner for a cylinder of 190806AB VSR, along with a fit that will be discussed below. 190806AB VSR was chosen for this work because its instantaneous modulus $G(0)$ is much larger than its static modulus $G(\infty)$ and its dynamics are slow, both of which make it easy to study.

The measured $G(t-t_{0})$ is approximately of the form
\begin{equation}
G(t-t_{0}) = A + B f(t-t_{0}),\label{eq:GForm}
\end{equation}
where the instantaneous modulus $G(0)$ observed immediately following the step in strain is $A+B$, the static modulus $G(\infty)$ observed long after the step in strain is $A$, and $f(t-t_{0})$ decreases monotonically from 1 to 0 as $t-t_{0}$ increases. The $B$ term is typical of an elastic solid while the $A f(t-t_{0})$ term is typical of a viscoelastic fluid.

To obtain a theoretical basis for Eq. \ref{eq:GForm}, along with an analytical expression for $f(t-t_{0})$, we assume that 190806AB VSR is well-described by the Fractional Zener model and study Eq. \ref{eq:FZMDE} for a step in strain $\epsilon_{0}$ at time $t_{0}$, 
\begin{align}
\epsilon(t)=\epsilon_{0}H(t-t_{0}),
\end{align}
where $H(x)$ is the Heaviside step function. For this $\epsilon(t)$, Eq. \ref{eq:FZMDE} becomes
\begin{multline}
\sigma(t)+\frac{F}{E_{2}}{}_{a}\textbf{D}^{\beta}_{t}\sigma(t)\\
=E_{1}\epsilon_{0}H(t-t_{0}) + F \left(\frac{E_{2}+E_{1}}{E_{2}}\right){}_{a}\textbf{D}^{\beta}_{t}\epsilon_{0}H(t-t_{0}),
\end{multline}
where $t \ge a$. Multiplying both sides by $E_{2}/F$ and defining
\begin{gather*}
\lambda_{1}=-\frac{E_{2}}{F}\\
A = \frac{E_{1}E_{2}\epsilon_{0}}{F}\\
B =  (E_{2}+E_{1})\epsilon_{0},
\end{gather*} 
gives
\begin{multline}
{}_{a}\textbf{D}^{\beta}_{t}\sigma(t)-\lambda_{1}\sigma(t)
=AH(t-t_{0}) +  B{}_{a}\textbf{D}^{\beta}_{t}H(t-t_{0}).\label{eq:FDEHSF}
\end{multline}

Equation \ref{eq:FDEHSF} is a non-homogeneous fractional differential equation of the type found in \cite{podlubny1998}, Example 4.3:
\begin{align}
{}_{a}\textbf{D}_{t}^{\alpha} y(t) - \lambda_{1} y(t) = h(t)\label{eq:Pod43}
\end{align}
for $t>a$. The solution to Eq. \ref{eq:Pod43}, for zero initial conditions at time $a$, is
\begin{align}
y(t)=\int_{a}^{t}(t-\tau)^{\alpha-1}E_{\alpha,\alpha}[\lambda_{1}(t-\tau)^{\alpha}]h(\tau)d\tau\label{eq:Pod43S}
\end{align}
Applying this result to Eq. \ref{eq:FDEHSF} gives
\begin{align}
\sigma(t) & =\int_{a}^{t}(t-\tau)^{\beta-1}E_{\beta,\beta}[\lambda_{1}(t-\tau)^{\beta}]\nonumber\\
& \hspace{1.0 cm}\times \left(AH(\tau-t_{0}) +  B{}_{a}\textbf{D}^{\beta}_{\tau}H(\tau-t_{0})\right)d\tau\nonumber\\
& = I_{1} + I_{2},\label{eq:I1I2}
\end{align}
where
\begin{gather}
I_{1} \equiv A\int_{a}^{t}(t-\tau)^{\beta-1}E_{\beta,\beta}[\lambda_{1}(t-\tau)^{\beta}] H(\tau-t_{0})d\tau\nonumber\\
I_{2} \equiv B\int_{a}^{t}(t-\tau)^{\beta-1}E_{\beta,\beta}[\lambda_{1}(t-\tau)^{\beta}]\nonumber {}_{a}\textbf{D}^{\beta}_{\tau}H(\tau-t_{0})d\tau
\end{gather}

To evaluate $I_{1}$, we use the Heaviside step function $H(x)$ to change the limits of integration, 
\begin{align}
I_{1} & = A\int_{t_{0}}^{t}(t-\tau)^{\beta-1}E_{\beta,\beta}[\lambda_{1}(t-\tau)^{\beta}]d\tau\label{eq:I11}
\end{align}
Substituting $z=(t-\tau)/(t-t_{0})$ in Eq. \ref{eq:I11} gives
\begin{align}
I_{1} & = A (t-t_{0})^{\beta}\int_{0}^{1}z^{\beta-1}E_{\beta,\beta}[\lambda_{1}(t-t_{0})^\beta z^{\beta}]dz\label{eq:I12}
\end{align}
Using Ref. \cite{mathai2008mittag} (2.2.14) to evaluate the integral in Eq. \ref{eq:I12} gives
\begin{align}
I_{1} & = A (t-t_{0})^{\beta}E_{\beta,\beta+1}[\lambda_{1} (t-t_{0})^\beta]\label{eq:I13}
\end{align}
Substituting $x=\lambda_{1} (t-t_{0})^{\beta}$ into Eq. \ref{eq:I13} and expanding $E_{\beta,\beta+1}[x]$ as its series gives
\begin{align}
I_{1}&=\frac{Ax}{\lambda_{1}}\sum_{k=0}^{\infty}\frac{x^{k}}{\beta k + \beta +1}\nonumber\\
&=\frac{A}{\lambda_{1}}\sum_{k=0}^{\infty}\frac{x^{k+1}}{\beta (k+1) +1}\nonumber\\
&=\left(\frac{A}{\lambda_{1}}\sum_{j=0}^{\infty}\frac{x^{j}}{\beta j +1}\right)-\frac{A}{\lambda_{1}}\nonumber\\
&=\frac{A}{\lambda_{1}}\left(E_{\beta}[x]-1\right)\label{eq:I14}
\end{align}
Restoring $x=\lambda_{1} (t-t_{0})^{\beta}$ in Eq. \ref{eq:I14} gives
\begin{align}
I_{1}=\frac{A}{\lambda_{1}}\left(E_{\beta}[\lambda_{1} (t-t_{0})^{\beta}]-1\right)\label{eq:I15}
\end{align}

To evaluate $I_{2}$, we use the definition ${}_{a}\textbf{D}_{t}^{\beta}H(x)$ from Eq. \ref{eq:RLLS}, followed by the Heaviside step function $H(x)$ to change the limits of integration,
\begin{align}
I_{2} & = B\int_{t_{0}}^{t}(t-\tau)^{\beta-1}E_{\beta,\beta}[\lambda_{1}(t-\tau)^{\beta}]\nonumber\\
& \hspace{1.0 cm}\times \frac{1}{\Gamma(1-\beta)}\frac{d}{d\tau}\int_{t_{0}}^{\tau}(\tau-\xi)^{-\beta}d\xi d\tau\\
& = B\int_{t_{0}}^{t}(t-\tau)^{\beta-1}E_{\beta,\beta}[\lambda_{1}(t-\tau)^{\beta}]\nonumber\\
& \hspace{1.0 cm}\times \frac{1}{\Gamma(1-\beta)}(\tau-t_{0})^{-\beta}d\tau\label{eq:I21}
\end{align}
Substituting $z=(t-\tau)/(t-t_{0})$ into Eq. \ref{eq:I21} gives
\begin{align}
I_{2} & =\frac{B}{\Gamma(1-\beta)}\int_{0}^{1}z^{\beta-1}(t-t_{0})^{\beta}
E_{\beta,\beta}[\lambda_{1}(t-t_{0})^{\beta}z^{\beta}]\nonumber\\
& \hspace{2.5 cm}\times (t-t_{0})^{-\beta}(1-z)^{-\beta}dz\nonumber\\
& = \frac{B}{\Gamma(1-\beta)}\int_{0}^{1}z^{\beta-1}
E_{\beta,\beta}[\lambda_{1}(t-t_{0})^{\beta}u^{\beta}](1-z)^{-\beta}dz\label{eq:I22}
\end{align}
Using Ref. \cite{mathai2008mittag} (2.2.14) to evaluate the integral in Eq. \ref{eq:I22} gives
\begin{align}
I_{2} & = \frac{B}{\Gamma(1-\beta)} \Gamma(1-\beta)E_{\beta}[\lambda_{1}(t-t_{0})^{\beta}]\nonumber\\
& = BE_{\beta}[\lambda_{1}(t-t_{0})^{\beta}]\label{eq:I23}
\end{align}

Combining Eqs. \ref{eq:I1I2}, \ref{eq:I15}, and \ref{eq:I23} and using the definitions of $\lambda_{1}$, $A$, and $B$ gives
\begin{align}
\sigma(t) &= \frac{A}{\lambda_{1}}\left(E_{\beta}[\lambda_{1} (t-t_{0})^{\beta}]-1\right)+BE_{\beta}[\lambda_{1}(t-t_{0})^{\beta}]\nonumber\\
&=-E_{1}\epsilon_{0}\left(E_{\beta}[-\frac{E_{2}}{F}(t-t_{0})^{\beta}]-1\right)+\nonumber\\
& \hspace{1.0 cm}(E_{1}+E_{2})E_{\beta}[-\frac{E_{2}}{F}(t-t_{0})^{\beta}]\\
&=E_{1}\epsilon_{0}+E_{2}\epsilon_{0}E_{\beta}[-\frac{E_{2}}{F}(t-t_{0})^{\beta}]\label{eq:I1I22}
\end{align}
It will be useful to define characteristic time $\tau_{1}$,
\begin{align}
\tau_{1}=\left(\frac{F}{E_{2}}\right)^{1/\beta},
\end{align} 
so that $\sigma(t)$ can be written
\begin{align}
\sigma(t)=E_{1}\epsilon_{0}+E_{2}\epsilon_{0}E_{\beta}\left[-\left(\frac{t-t_{0}}{\tau_{1}}^{\beta}\right)\right].\label{eq:SRSol}
\end{align}

Dividing this stress $\sigma(t)$ by the step in strain $\epsilon_{0}$ at time $t_{0}$ that caused it gives the Fractional Zener model's stress relaxation modulus $G_{\textsc{fz}}(t-t_{0})$,
\begin{align}
G_{\textsc{fz}}(t-t_{0})=E_{1}+E_{2}E_{\beta}\left[-\left(\frac{t-t_{0}}{\tau_{1}}^{\beta}\right)\right].\label{eq:GFZ}
\end{align}
The fit shown in Fig. \ref{fig:Compression} was made using \ref{eq:GFZ} and the given values for $E_{1}$, $E_{2}$, $\beta$, and $\tau_{1}$ where obtained from that fit. We note that $G_{\textsc{fz}}(t-t_{0})$ does indeed have the form anticipated in Eq. \ref{eq:GForm}.

\subsection{Shape Recovery}
\begin{figure}
	\includegraphics[width=0.8\columnwidth]{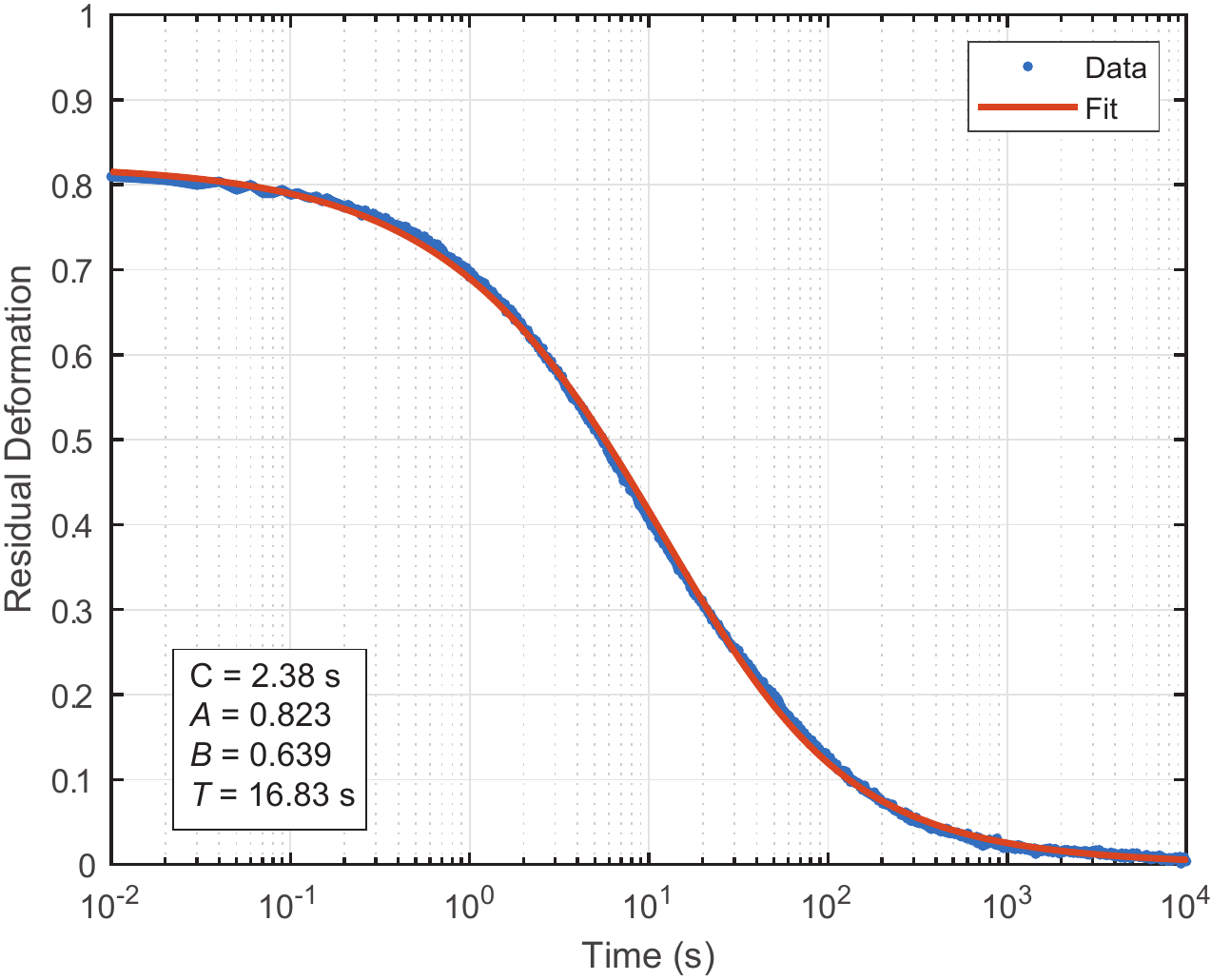}\\%
	\includegraphics[width=0.8\columnwidth]{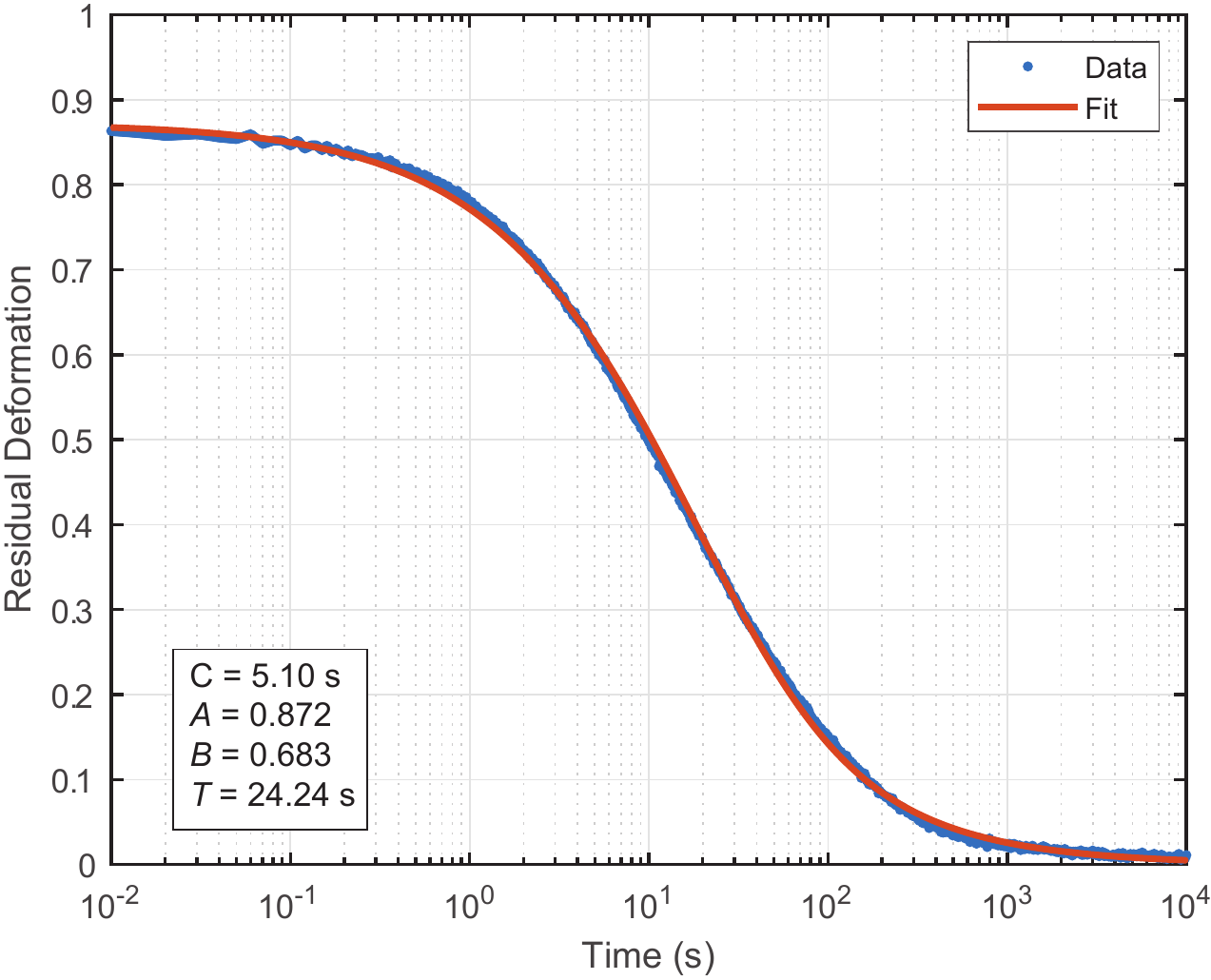}\\%
	\includegraphics[width=0.8\columnwidth]{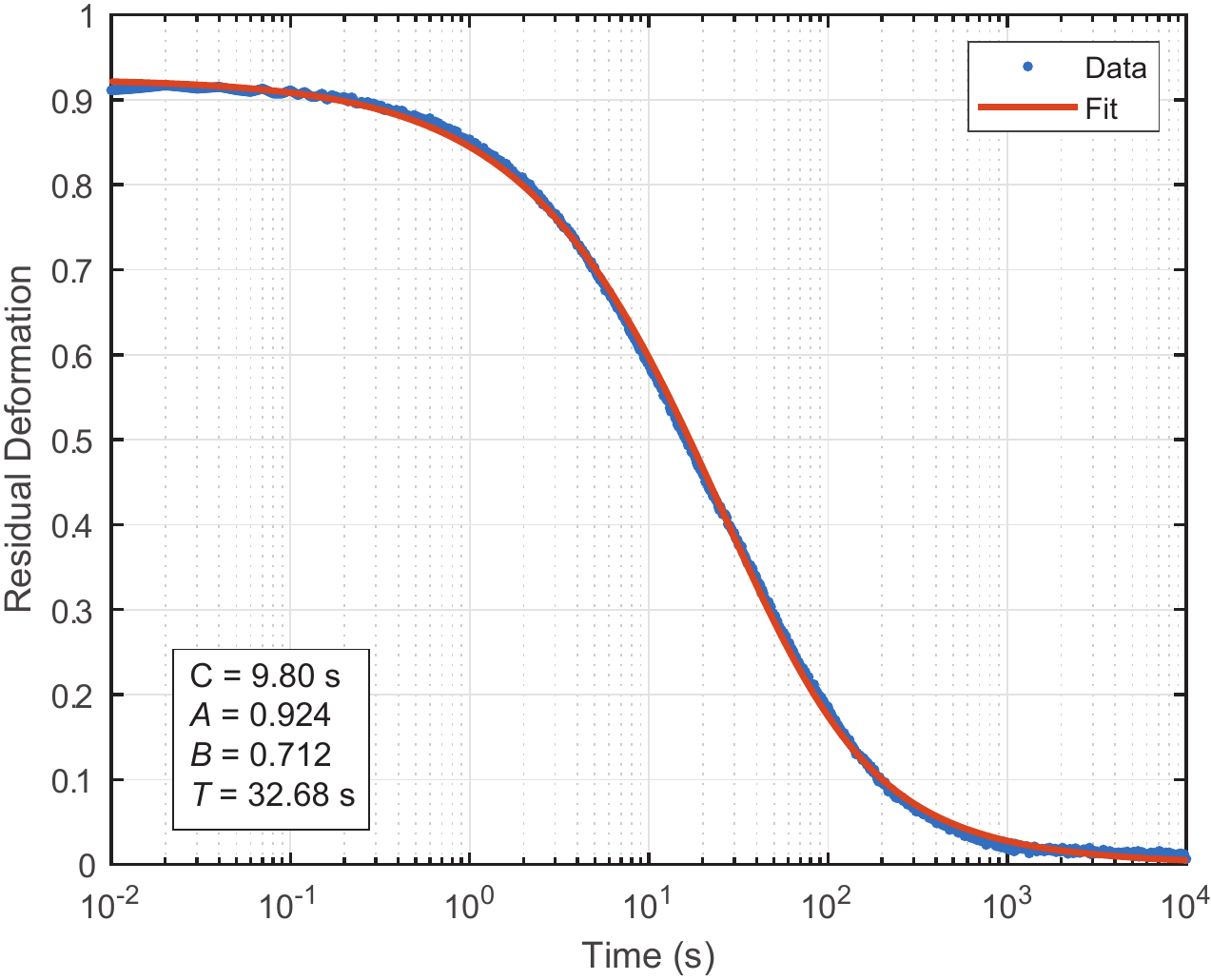}%
	\caption{Shape recovery of a VSR cylinder after a unit compression of duration C. The data are fit by Eq. \ref{eq:SRApprox} for $\epsilon_{0}=1$. Instantaneous residual deformation $A$, shape-recovery fractional order $B$, shape-recovery characteristic time $T$ are the equation's three parameters.\label{fig:S2to10}}
\end{figure}

\begin{figure}
	\includegraphics[width=0.8\columnwidth]{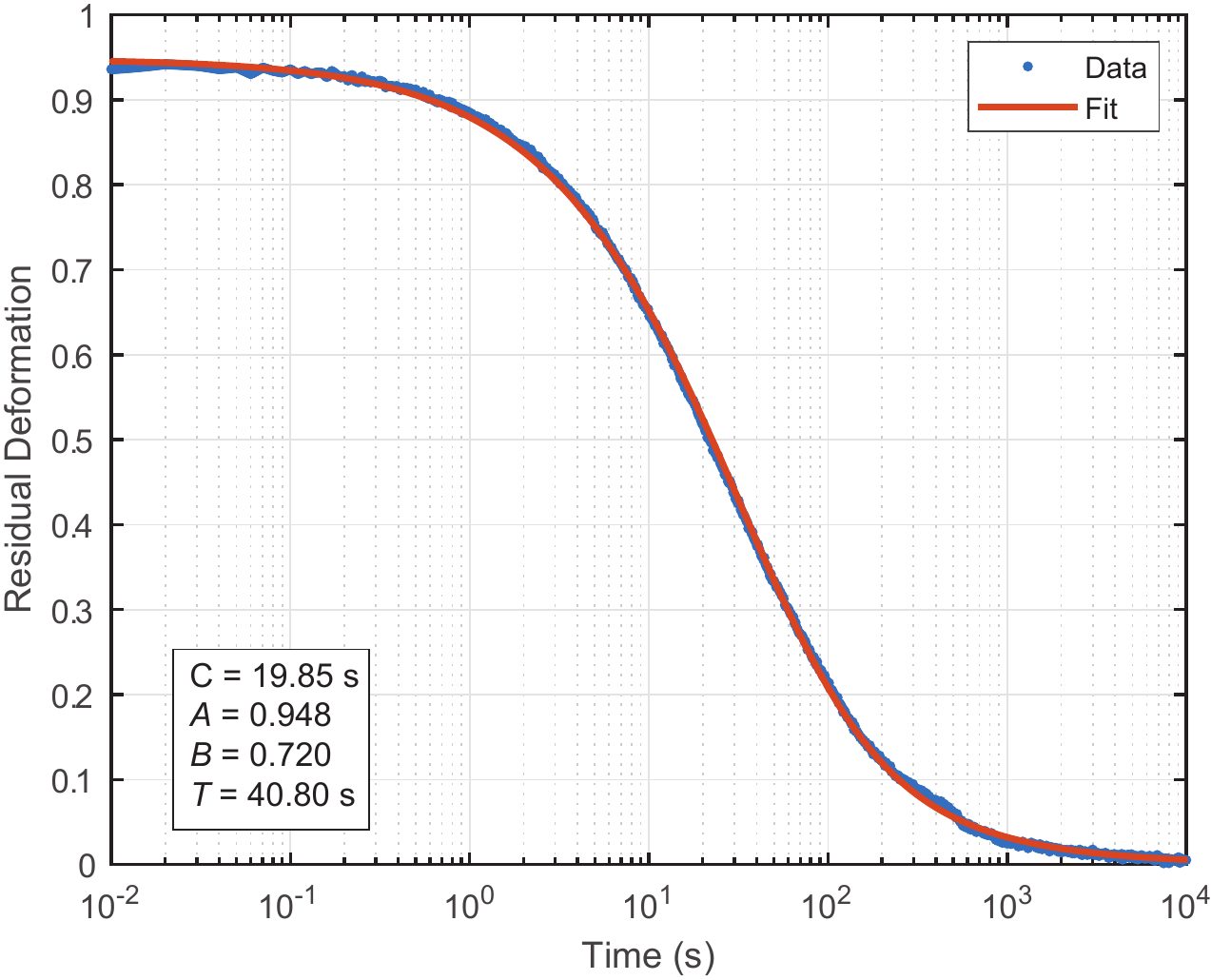}\\%
	\includegraphics[width=0.8\columnwidth]{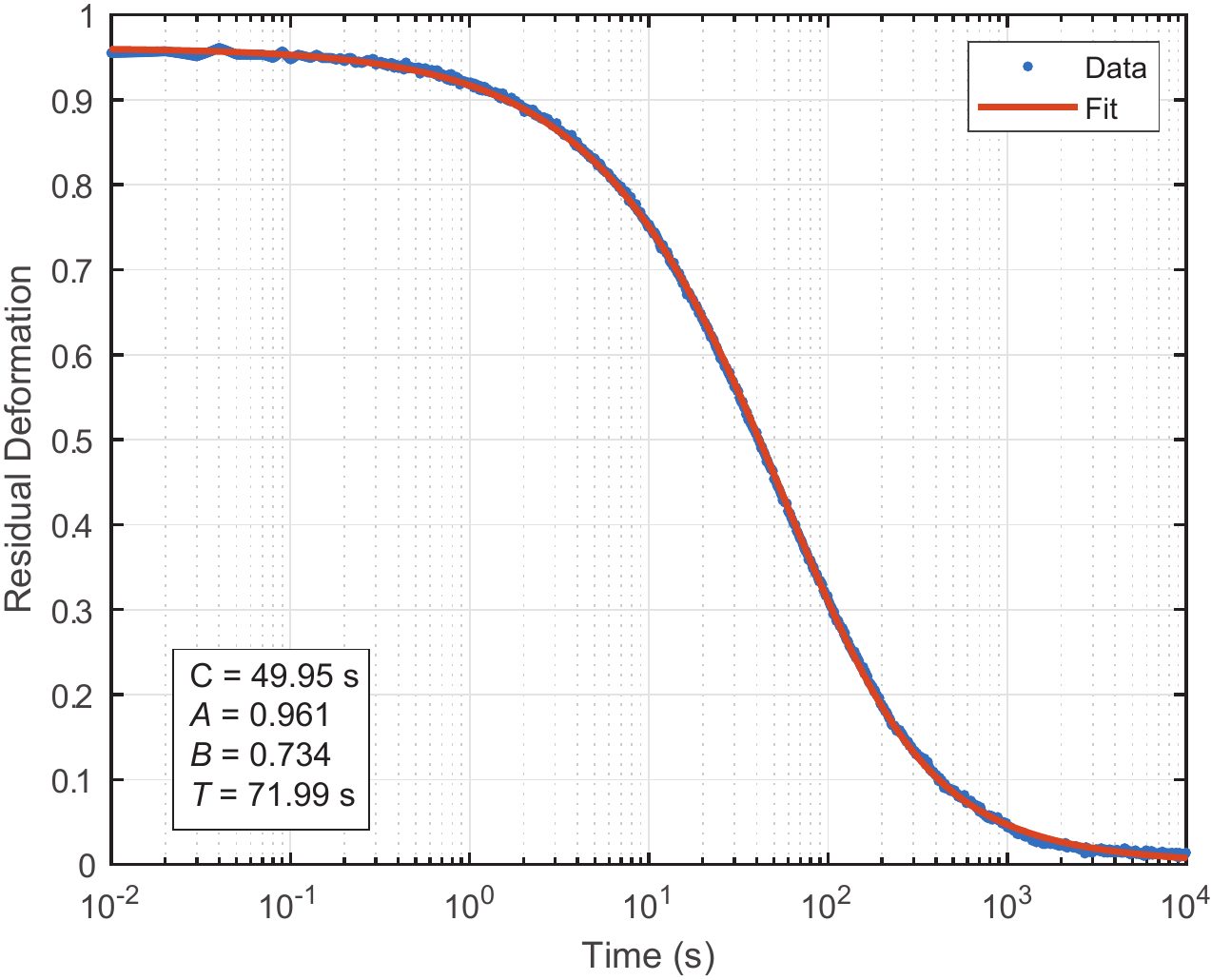}\\%
	\includegraphics[width=0.8\columnwidth]{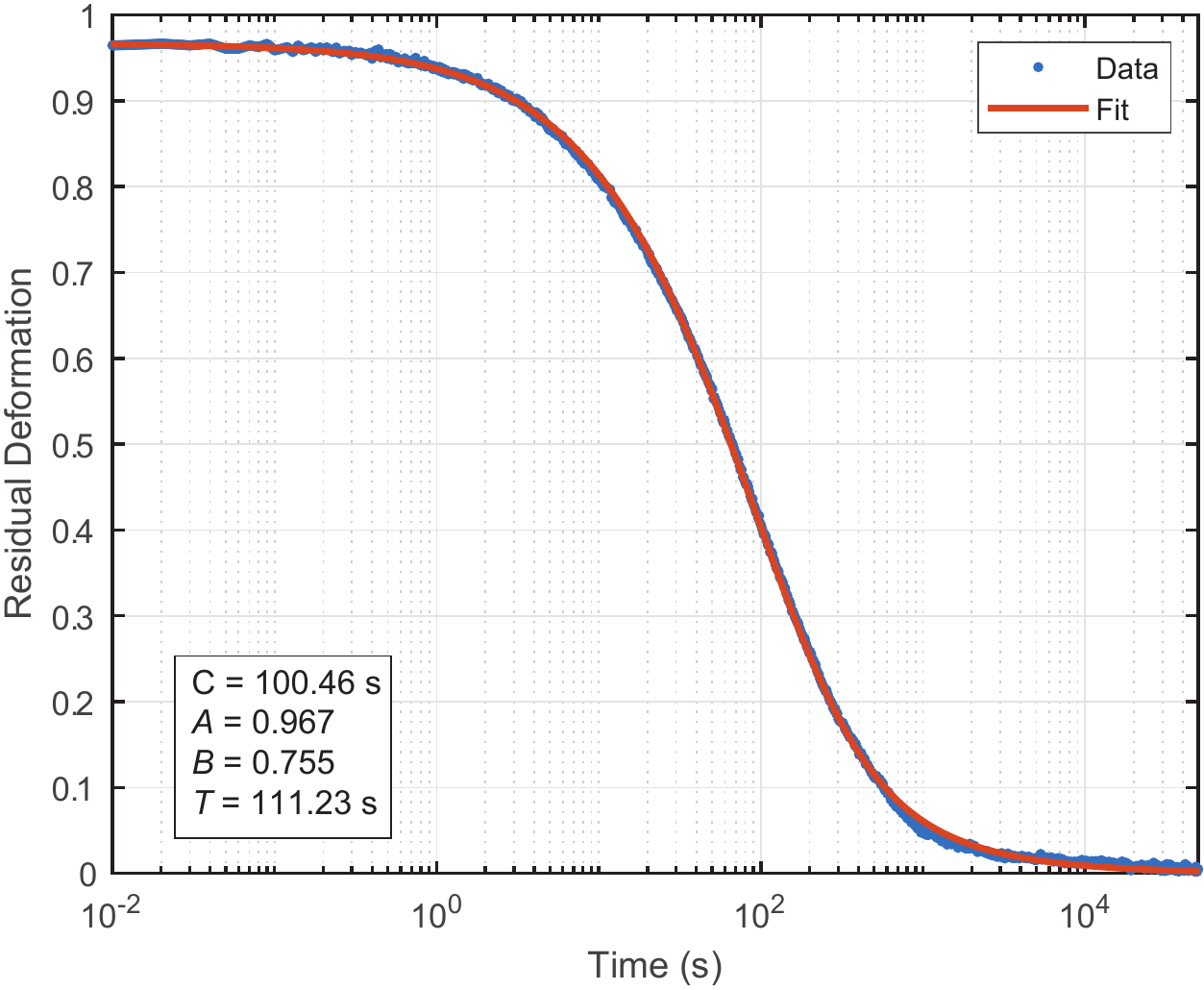}%
	\caption{Same as Fig. \ref{fig:S2to10}.\label{fig:S20to100}}
\end{figure}

\begin{figure}
	\includegraphics[width=0.8\columnwidth]{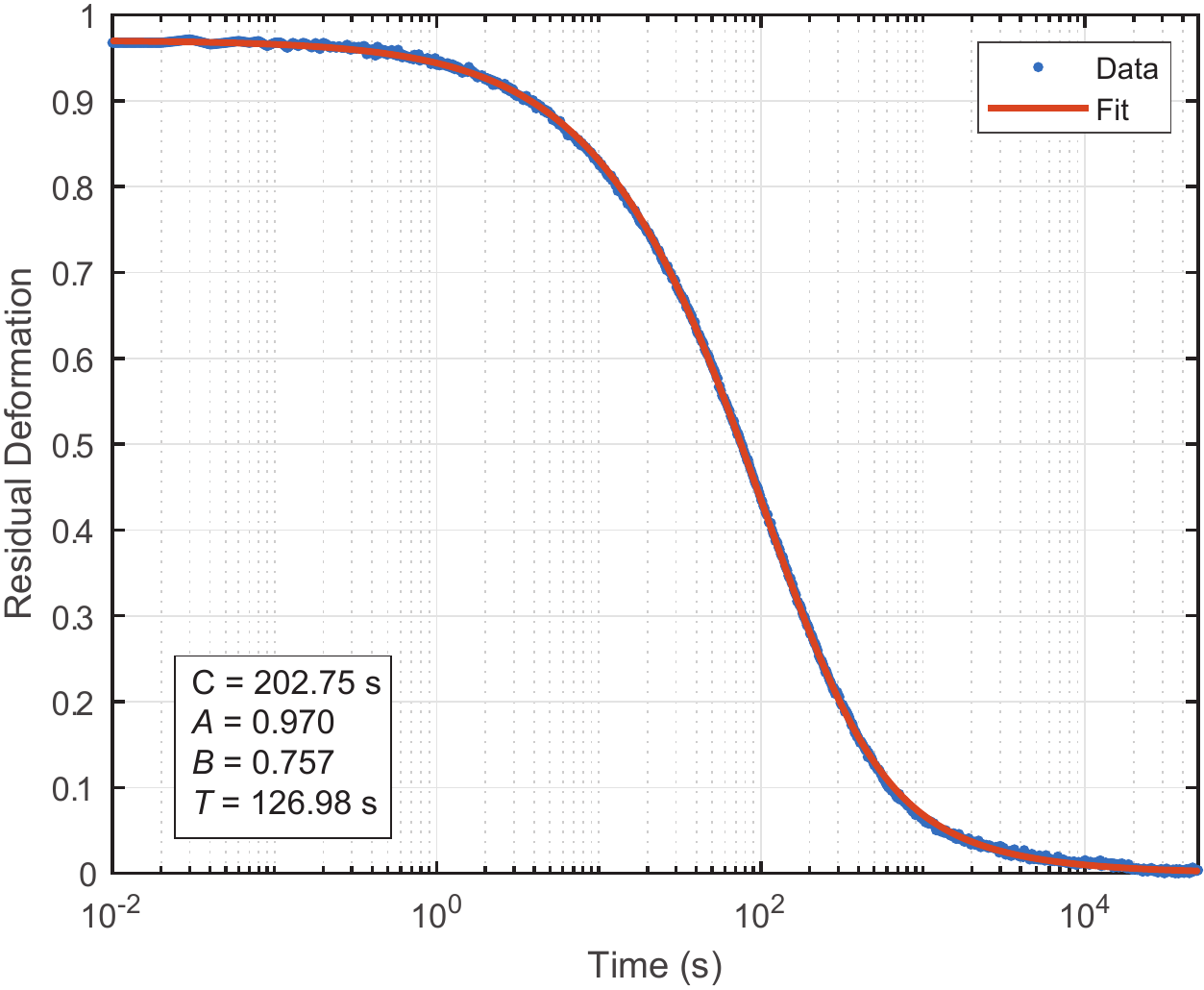}\\%
	\includegraphics[width=0.8\columnwidth]{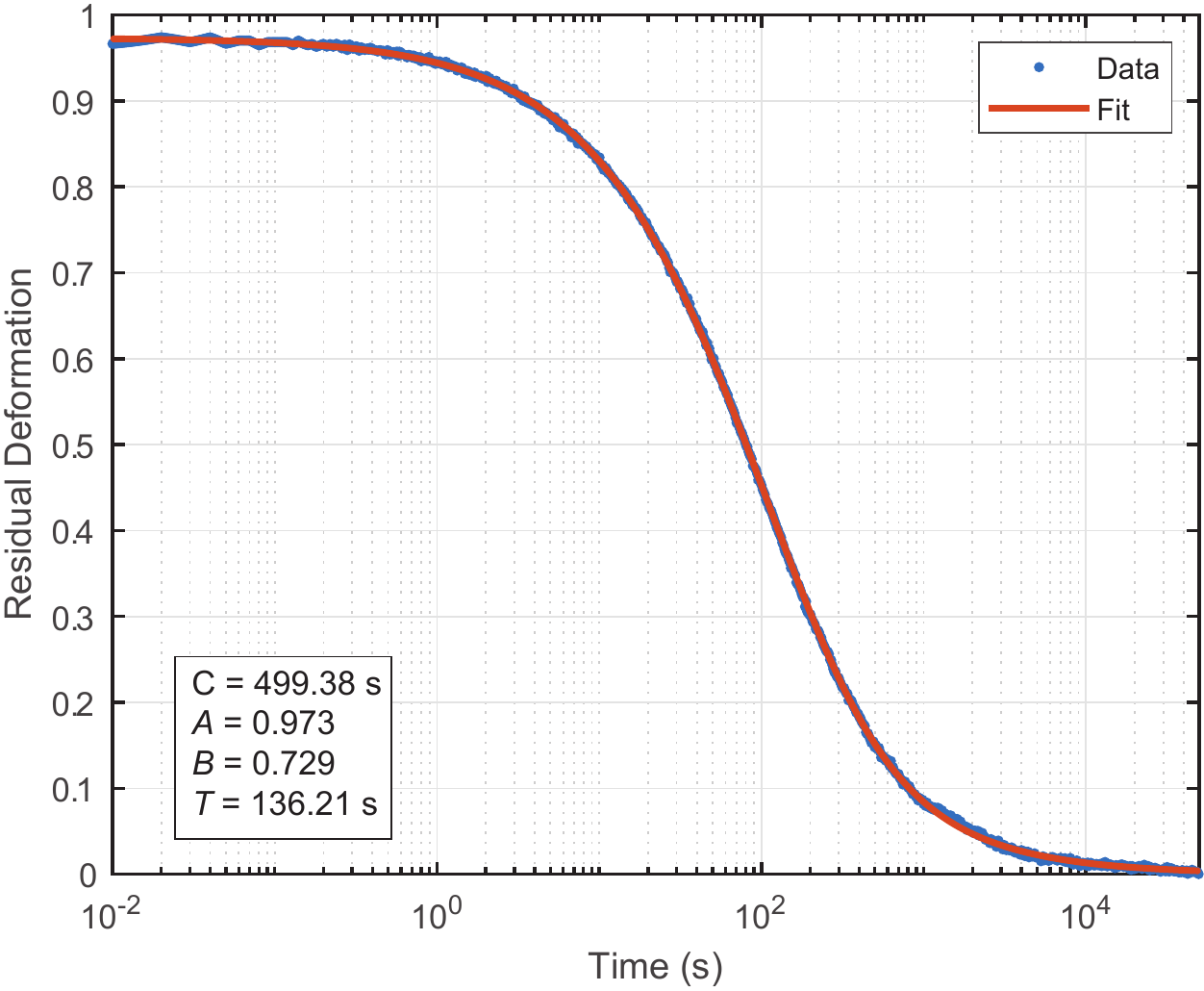}\\%
	\includegraphics[width=0.8\columnwidth]{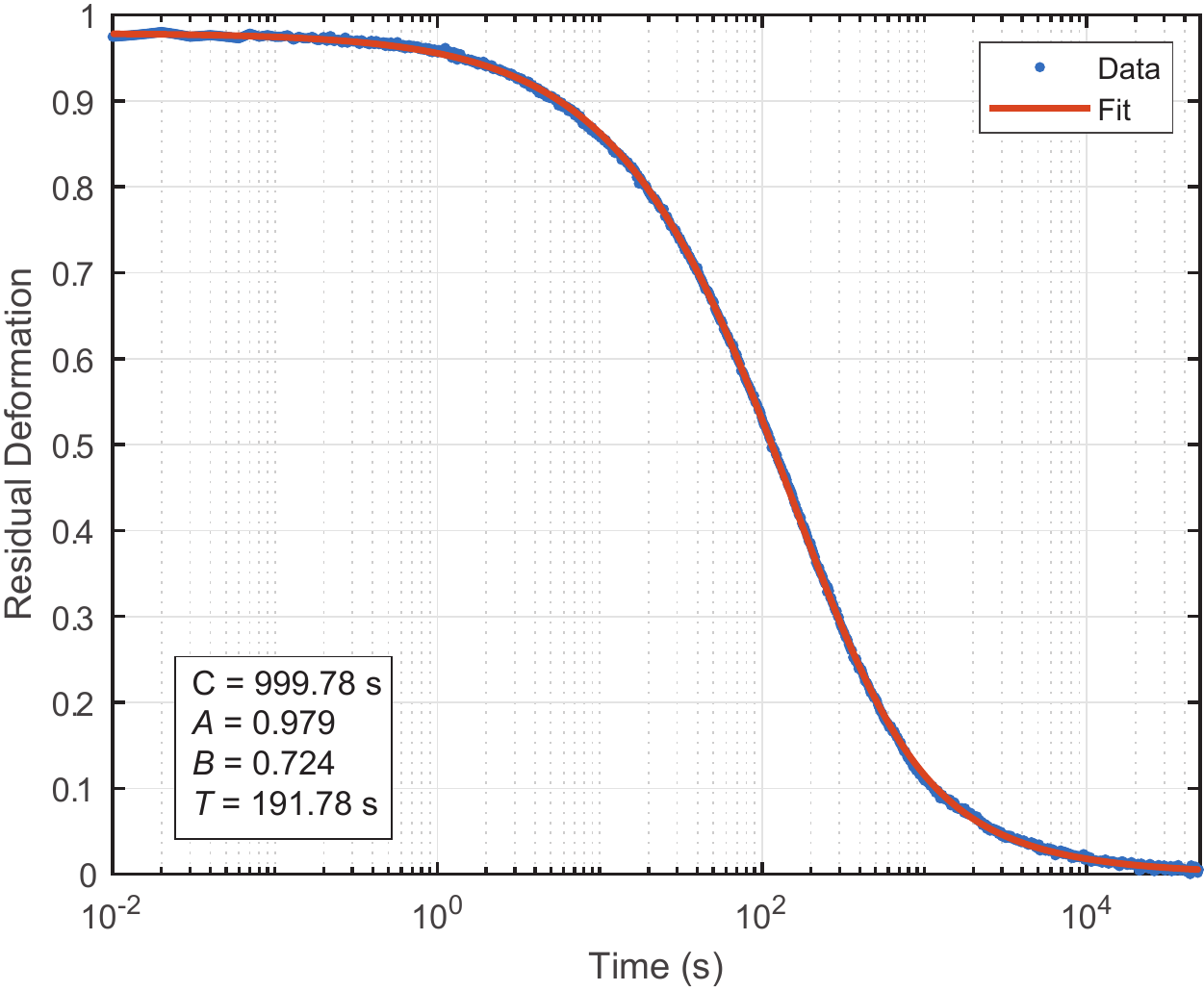}%
	\caption{Same as Fig. \ref{fig:S2to10}.\label{fig:S200to1000}}
\end{figure}

To study shape recovery, a fully relaxed 9.53mm-diameter cylinder of VSR is compressed suddenly from 6.35mm tall to 4.86mm tall, held at constant strain for a period of time, and then released suddenly from external stress. As the cylinder gradually recovers its original shape, its strain is measured at 10ms intervals by a LVDT. Figs. \ref{fig:S2to10}, \ref{fig:S20to100}, and \ref{fig:S200to1000} show measurements obtained for 9 different compression times, ranging from 2.38 s to 999.78 s, along with fits to the data that will be discussed below.


This shape-recovery study is essentially an interrupted version of the stress-relaxation study. As before, there is a step in strain $\epsilon_{0}$ at time $t_{0}$ and that strain continues until time $t_{1}$. At that moment, the stress is abruptly decreased to zero and the two studies begin to differ. Because the studies are so similar, however, the mathematical approach and fractional differential equation (Eq. \ref{eq:FZMDE}) used to model the compression study can also be used to model the shape-recovery study. For the shape-recovery study, however, the goal is to obtain $\epsilon(t)$ during the period $t \ge t_{1}$.

Because the two studies are identical up until time $t_{1}$, $\sigma(t)$ for the shape-recovery study is a truncated version of Eq. \ref{eq:SRSol},
\begin{align}
\sigma(t) & =\left(E_{1}\epsilon_{0}+E_{2}\epsilon_{0}E_{\beta}\left[-\left(\frac{t-t_{0}}{\tau_{1}}^{\beta}\right)\right]\right)\nonumber\\
& \hspace{1.0 cm}\times H(t-t_{0})(1-H(t-t_{1})),\label{eq:SigmaHeavi}
\end{align}
where the Heaviside step functions $H(t)$ reduce $\sigma(t)$ to zero for $t<t_{0}$ and $t>t_{1}$.

With stress $\sigma(t)$ known at all times and strain $\epsilon(t)$ known for $t < t_{1}$, solving Eq. \ref{eq:FZMDE} for $\epsilon(t)$ at $t \ge t_{1}$ is all that needs to be done. Unfortunately, solving Eq. \ref{eq:FZMDE} analytically in this more-complicated circumstance is beyond our abilities.

Ordinary integer time derivatives are local in time, meaning that (finite) integer-order time derivatives of $f(t)$ depend only on $f(t')$ at times $t'$ infinitesimally close to $t$. In contrast, fractional time derivatives are inherently non-local, meaning that non-integer fractional-order time derivatives of $f(t)$ depend on $f(t')$ at times $t'$ both infinitesimally close to $t$ and throughout the past ($t' < t$). In other words, fractional time derivatives have memory and the operators that perform them examine the entire pasts of the functions they operate on.

In Section C below, we develop the analytic solution for strain $\epsilon(t)$ at $t \ge t_{1}$ up until its integrals and derivatives of integrals become nearly intractable and we are forced to admit defeat. Rather than walking away, however, we turn to approximations. Specifically, we can find an approximate solution for $\epsilon(t)$ at $t \ge t_{1}$ by treating the compression and shape recovery periods separately, as though the compression period merely set the stage for the shape recovery period. Treating the shape recovery period separately moves the lower terminal $a$ of the fractional derivatives in Eq. \ref{eq:FZMDE} to time $t_{1}$ and thereby discards their memory of the compression period.

Since stress $\sigma(t)$ is zero during the shape recovery period, its memory-truncated fractional derivative ${}_{t_{1}}\textbf{D}_{t}^{\beta}\sigma(t)$ is also zero. The reduced version of Eq. \ref{eq:FZMDE}, without $\sigma(t)$ and starting at time $t_{1}$, is
\begin{align}
E_{1}\epsilon(t) + F \left(\frac{E_{2}+E_{1}}{E_{2}}\right){}_{t_{1}}\textbf{D}^{\beta}_{t}\epsilon(t) = 0,
\end{align}
for $t \ge t_{1}$. Defining
\begin{align}
\lambda_{2}=-\frac{E_{1}E_{2}}{F(E_{1}+E_{2})},
\end{align}
and rearranging gives
\begin{align}
{}_{t_{1}}\textbf{D}^{\beta}_{t}\epsilon(t)=\lambda_{2}\epsilon(t),\label{eq:SRFDE}
\end{align}
The solution to Eq. \ref{eq:SRFDE}, obtained below in Section A, is
\begin{align}
\epsilon(t)=\epsilon(t_{1})E_{\beta}\left[\lambda_{2}(t-t_{1})^{\beta}\right],\label{eq:SRGen}
\end{align}
where $\epsilon(t_{1})$ is the initial value of $\epsilon(t)$. That initial strain $\epsilon(t_{1})$ depends on what happened during the compression period and is, in fact, the only recollection of the compression period that survives in this memory-truncated approximation of the shape-recovery process.

To determine $\epsilon(t_{1})$, we consider the transition from the compression period to the shape recovery period. That transition truncates the memory in the fractional derivatives and the two springs have no memory at all, however, one physical quantity survives the transition: the spring-pot's strain $\epsilon_{3}$. In fact, for the spring-pot's stress $\sigma_{2}(t)$ to remain finite, Eq. \ref{eq:ESpringpot} requires that the spring-pot's strain $\epsilon_{3}(t)$ be continuous. Thus $\epsilon_{3}(t_{1})$ must have the same value at the beginning of the shape recovery period as it had at the end of the compression period.

At the end of the compression period, the strain is $\epsilon_{0}$ and the stress is
\begin{align}
\sigma(t_{1}) & = E_{1}\epsilon_{0}+E_{2}\epsilon_{0}E_{\beta}\left[-\left(\frac{t_{1}-t_{0}}{\tau_{1}}^{\beta}\right)\right]\nonumber\\
& = \sigma_{1}(t_{1})+\sigma_{2}(t_{1}).
\end{align}
It is easy to show that
\begin{align}
\sigma_{2}(t_{1})=E_{2}\epsilon_{0}E_{\beta}\left[-\left(\frac{t_{1}-t_{0}}{\tau_{1}}^{\beta}\right)\right]
\end{align}
Equations \ref{eq:SE1} and \ref{eq:SE4} can then be used to determine the spring-pot's strain $\epsilon_{3}(t)$ at time $t_{1}$,
\begin{align}
\epsilon_{3}(t_{1}) & = \epsilon_{0}-\frac{1}{E_{2}}\sigma_{2}(t_{1})\nonumber\\
& = \epsilon_{0}-\frac{1}{E_{2}}E_{2}\epsilon_{0}E_{\beta}\left[-\left(\frac{t_{1}-t_{0}}{\tau_{1}}^{\beta}\right)\right]\nonumber\\
& = \epsilon_{0}\left(1-E_{\beta}\left[-\left(\frac{t_{1}-t_{0}}{\tau_{1}}^{\beta}\right)\right]\right).\label{eq:e3Com}
\end{align}

To avoid a divergence of stress, $\epsilon_{3}(t_{1})$ at the start of the shape recovery period must equal $\epsilon_{3}(t_{1})$ at the end of the compression period, specifically Eq. \ref{eq:e3Com}. During the shape-recovery period, $\epsilon_{3}(t_{1})$ can be used to find $\epsilon(t)$ using the relationship
\begin{align}
\epsilon(t_{1})=\frac{E_{2}}{E_{1}+E_{2}}\epsilon_{3}(t_{1}),
\end{align}
obtained using Eqs. \ref{eq:SE1}--\ref{eq:SE4} along with $\sigma(t)=0$. Thus
\begin{align}
\epsilon(t_{1})=\frac{E_{2}\epsilon_{0}}{E_{1}+E_{2}}\left(1-E_{\beta}\left[-\left(\frac{t_{1}-t_{0}}{\tau_{1}}^{\beta}\right)\right]\right).
\end{align}
Using this initial value in Eq. \ref{eq:SRGen} gives the strain $\epsilon(t)$ for the shape recovery period,
\begin{align}
\epsilon(t) & =\frac{E_{2}\epsilon_{0}}{E_{1}+E_{2}}\left(1-E_{\beta}\left[-\left(\frac{t_{1}-t_{0}}{\tau_{1}}^{\beta}\right)\right]\right)\nonumber\\
& \hspace{1.0 cm}\times E_{\beta}\left[\lambda_{2}(t-t_{1})^{\beta}\right].\label{eq:SRFinal}
\end{align}

It will be useful to define
\begin{align}
\tau_{2}=-\lambda_{2}^{-1/\beta}=\left(\frac{F(E_{1}+E_{2})}{E_{1}E_{2}}\right)^{1/\beta}\\
= \left(\frac{E_{1}+E_{2}}{E_{1}}\right)^{1/\beta}\tau_{1},\label{eq:Tau2fromTau1}
\end{align} 
so that the solution can be written
\begin{align}
\epsilon(t) & =\frac{E_{2}\epsilon_{0}}{E_{1}+E_{2}}\left(1-E_{\beta}\left[-\left(\frac{t_{1}-t_{0}}{\tau_{1}}^{\beta}\right)\right]\right)\nonumber\\
& \hspace{1.0 cm}\times E_{\beta}\left[-\left(\frac{t-t_{1}}{\tau_{2}}\right)^{\beta}\right].\label{eq:SRFinalTau2}
\end{align}

Because the fractional derivatives' memory before time $t_{1}$ has been discarded, Eq. \ref{eq:SRFinalTau2} is only a rough approximation. For short compression periods, time $t_{0}$ is in the recent past and discarding its memory surely makes the approximation poor. Where the approximation should most closely resemble the actual solution for $\epsilon(t)$ is when the compression period is relatively long ($t_{1}-t_{0} \gg \tau_{1}$) and the stress $\sigma(t)$ is essentially constant as time $t_{1}$ approaches.

Although Eq. \ref{eq:SRFinalTau2} is an approximation, it suggests an expression with which to fit the experimental data in Figs. \ref{fig:S2to10}-\ref{fig:S200to1000}. If we redefine $t$ as the time since $t_{1}$ (the start of shape recovery) and consider $\epsilon(t)$ after an infinite compression ($t_{1}-t_{0} \rightarrow \infty$), Eq. \ref{eq:SRFinalTau2} simplifies to
\begin{align}
\epsilon(t)=\frac{E_{2}\epsilon_{0}}{E_{1}+E_{2}}E_{\beta}\left[-\left(\frac{t}{\tau_{2}}\right)^{\beta}\right].\label{eq:SRInfinite}
\end{align}
For a finite compression, a reasonable guess for $\epsilon(t)$ takes the same form as Eq. \ref{eq:SRInfinite},
\begin{align}
\epsilon(t)=A\epsilon_{0}E_{B}\left[-\left(\frac{t}{T}\right)^{B}\right],\label{eq:SRApprox}
\end{align}
but with parameters $A$, $B$, and $T$ that can be interpreted as follows: $A$ is the fraction of compressive strain that the system retains the moment after stress is released, $B$ is the fractional order of the shape-recovery process and likely to be related to the system's fractional order $\beta$, and $T$ is the characteristic time of the system's shape-recovery process and likely to be related to the system's characteristic times $\tau_{1}$ and $\tau_{2}$.

Fits of Eq. \ref{eq:SRApprox} to the VSR measurement data in Figs. \ref{fig:S2to10}-\ref{fig:S200to1000} are shown in those figures, along with values of $A$, $B$, and $T$ obtained from those fits. The fits are excellent, indicating that Eq. \ref{eq:SRApprox} has approximately the right form to describe the shape-recovery processes in both VSRs and the Fractional Zener model. The fit values for $A$, $B$, and $T$ are assembled together in Table \ref{tab:SRTable}.

\begin{table}
	\caption{Parameter values obtained by fitting Eq. \ref{eq:SRApprox} to the shape-recovery measurement data of Figs. \ref{fig:S2to10}-\ref{fig:S200to1000}. $A_{0}$ is the predicted fraction of compressive strain initially retained by the VSR, $A$ is the measured value, $B$ is the fractional order of the shape-recovery process, and $T$ is the characteristic time of the shape-recovery process. Uncertainties (95\% confidence) are approximately $A: \pm0.003$, $B: \pm0.002$, and $T: \pm 0.4$ s, based on the fits, however, $T$ is sensitive to atmospheric moisture. These data were taken at approximately 46\% RH, but $T$ can vary by 20\% or more with the air's humidity.\label{tab:SRTable}}
	\begin{ruledtabular}
		\begin{tabular}{rrrrr}
			\textrm{Compression (s)}&
			\textrm{$A_{0}$}&
			\textrm{$A$}&
			\textrm{$B$}&
			\textrm{$T$ (s)}\\
			\colrule
			2.38 & 0.838 & 0.823 & 0.639 & 16.83\\
			5.10 & 0.903 & 0.872 & 0.683 & 24.24\\
			9.80 & 0.935 & 0.924 & 0.712 & 32.68\\
			19.85 & 0.954 & 0.948& 0.720 & 40.80\\
			49.95 & 0.968 & 0.961 & 0.734 & 71.99\\
			100.46 & 0.974 & 0.967 & 0.755 & 111.23\\
			202.75 & 0.977 & 0.970 & 0.757 & 126.98\\
			499.38 & 0.980 & 0.973 & 0.729 & 136.21\\
			999.78 & 0.981 & 0.979 & 0.724 & 191.78\\
		\end{tabular}
	\end{ruledtabular}
\end{table}

Also shown in Table \ref{tab:SRTable} is $A_{0}$, the predicted fraction of compressive strain that the VSR retains the moment after stress is released. The values shown were calculated by dividing the first term of Eq. \ref{eq:SRFinalTau2} by $\epsilon_{0}$ to give
\begin{align}
A_{0}(t_{1}-t_{0})=\frac{E_{2}}{E_{1}+E_{2}}\left(1-E_{\beta}\left[-\left(\frac{t_{1}-t_{0}}{\tau_{1}}^{\beta}\right)\right]\right),
\end{align}
and using values of $E_{1}$, $E_{2}$, $\beta$, and $\tau_{1}$ obtained by the compression measurement (Fig. \ref{fig:Compression}).
The predicted values $A_{0}$ and measured values $A$ are quite similar, confirming the expectations that a VSR's strain rebound immediately following the release of stress is well-described by the Fractional Zener model and that the spring-pot's strain $\epsilon_{3}(t)$ does not change between the end of the compression period and the start of the shape-recovery period. 

\begin{figure}
	\includegraphics[width=0.8\columnwidth]{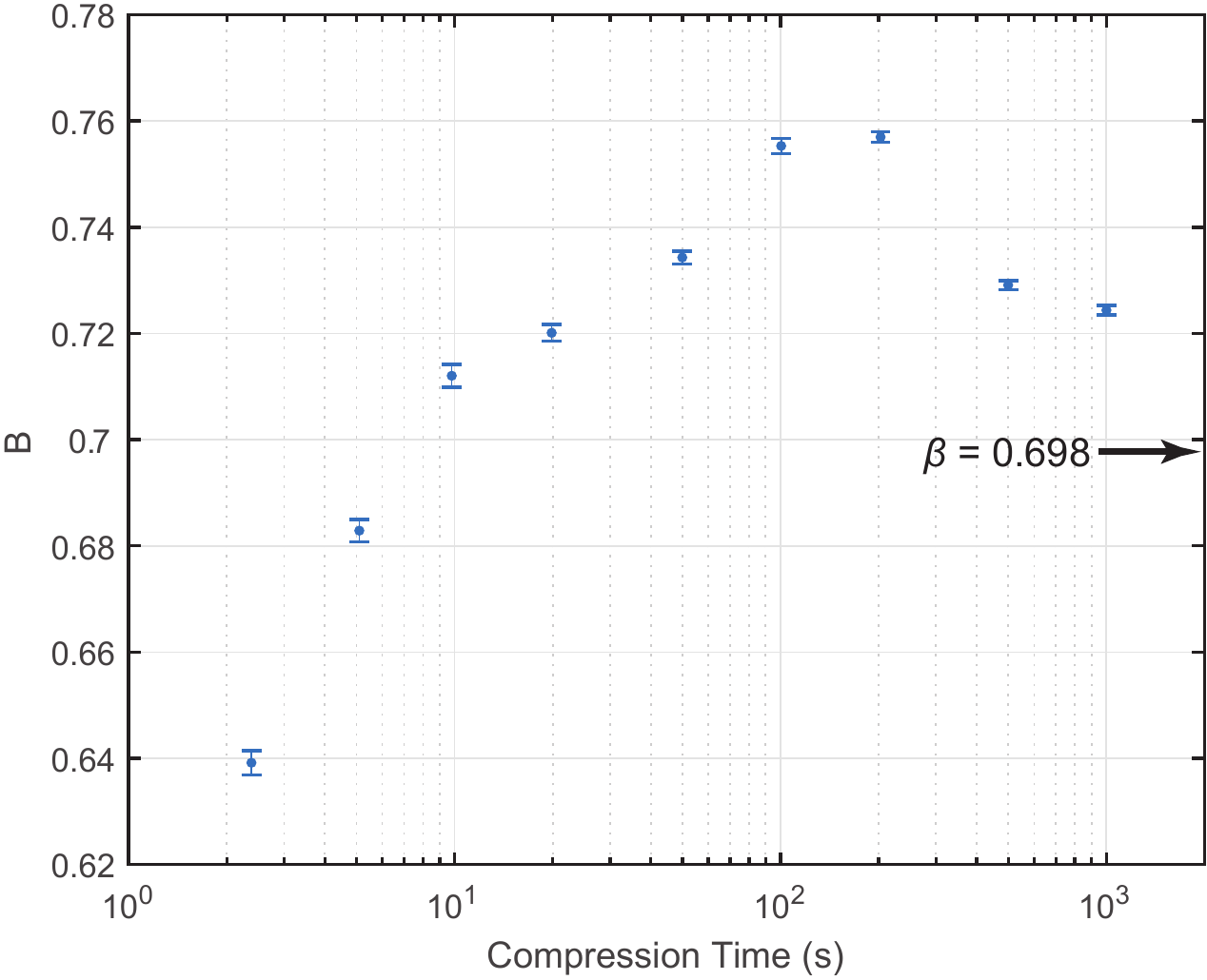}%
	\caption{Fractional order $B$ of the shape-recovery process as a function of compression duration. Fractional order $\beta$, obtained by fitting the Fractional Zener model to the compression measurement of Fig. \ref{fig:Compression}, is also shown. 95\% confidence intervals are those of the fits in Figs. \ref{fig:S2to10}--\ref{fig:S200to1000}.\label{fig:figBeta}}
\end{figure}

Figure \ref{fig:figBeta} shows the measured shape-recovery fractional-order $B$ as a function of compression duration. Also shown is the fractional order $\beta$, obtained by fitting the Fractional Zener model to the compression measurement of Fig. \ref{fig:Compression}. Comparing Eqs. \ref{eq:SRInfinite} and \ref{eq:SRApprox}, it seems likely that $B\rightarrow \beta$ as $t_{1}-t_{0}\rightarrow \infty$. We observe that $B$ starts relatively small at short compression times, rises to a peak value at compression times between 100 and 200 seconds, then decreases approximately toward $\beta$ as the compression time continues to increase. 

\begin{figure}
	\includegraphics[width=0.8\columnwidth]{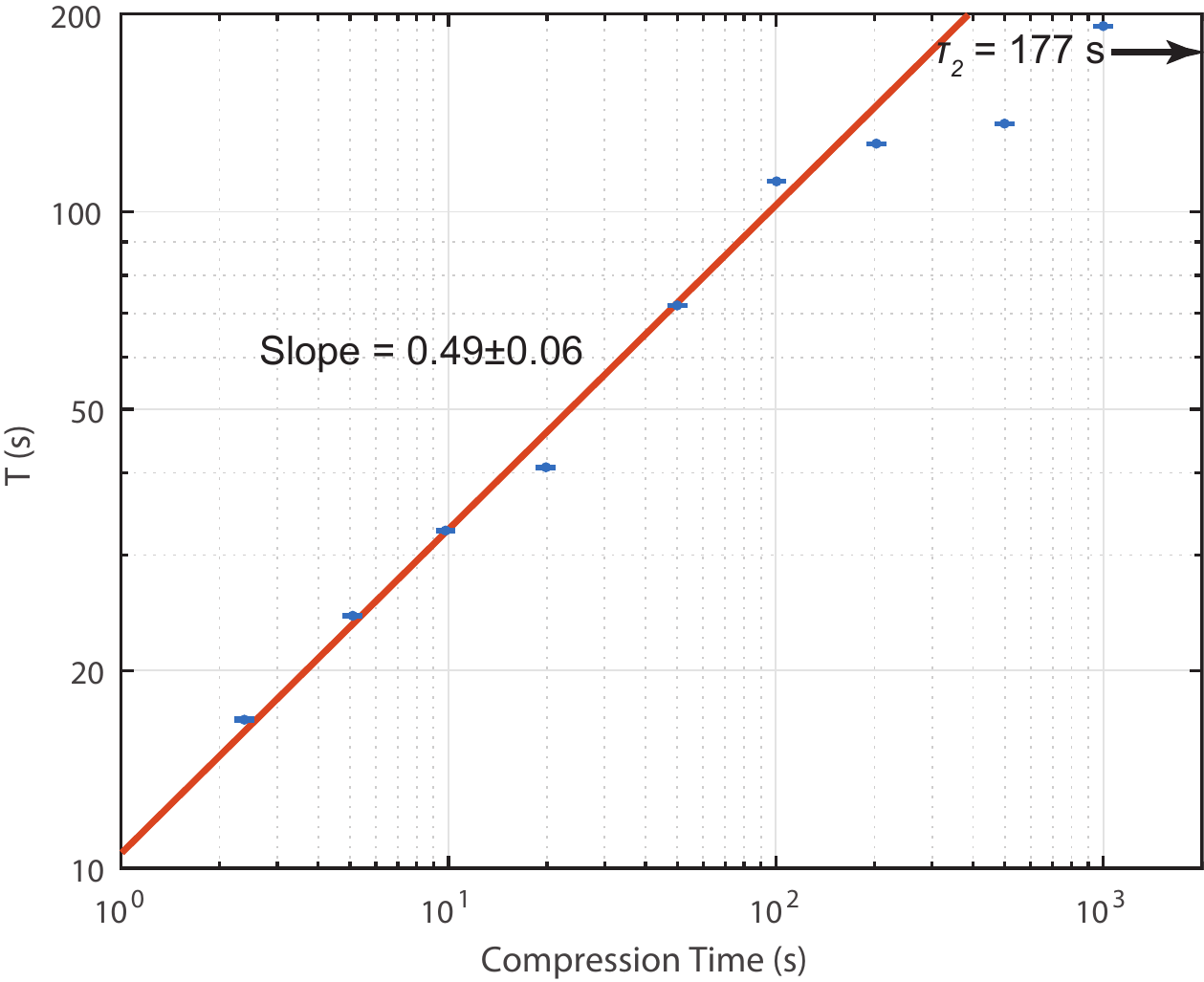}%
	\caption{Chacteristic time $T$ of the shape-recovery process as a function of compression duration. Shape-recovery characteristic time $\tau_{2}$, calculated by applying Eq. \ref{eq:Tau2fromTau1} to values obtained by fitting the Fractional Zener model to the compression measurement of Fig. \ref{fig:Compression}, is also shown. 95\% confidence intervals are those in the fits of Figs. \ref{fig:S2to10}--\ref{fig:S200to1000}, which were taken in approximately 46\% RH air. Note that $T$ is quite sensitive to atmospheric moisture and can vary by 20\% or more with humidity.\label{fig:figTau}}
\end{figure}

Figure \ref{fig:figTau} shows the measured shape-recovery characteristic time $T$ as a function of compression duration. Also shown is the 177 s value for $\tau_{2}$, calculated from Eq. \ref{eq:Tau2fromTau1} using value of $E_{1}$, $E_{2}$, $\beta$, and $\tau_{1}$ obtained by fitting the Fractional Zener model to the compression measurement of Fig. \ref{fig:Compression}. We observe that $T$ starts small at short compression times and rises linearly on the log-log scale until compression times longer than 100 s. A line fit to the shorter compression times has a slope of $0.49\pm 0.06$, suggesting that $T \propto \textrm{compression time}^{1/2}$.

At compression times longer than 100 s, $T$ appears to roll off. The value of $T$ for an infinite compression time cannot be determined experimentally, but it is likely to be in the vicinity of $\tau_{2}$. Unfortunately, the $\tau_{1}$ and $T$ values measured in VSRs are quite sensitive to atmospheric moisture and relative humidity, so more complete and careful measurements are unlikely to give additional insights.

\section{Conclusions}

The behaviors of viscoelastic silicone rubber are well-described by the Fractional Zener viscoelastic model. As shown previously,\cite{bloomfield2018} VSR's measured stress relaxation modulus $G(t-t_{0})$ is well-fit by the Fractional Zener model's stress relaxation modulus $G_{\textsc{fz}}(t-t_{0})$, given in Eq. \ref{eq:GFZ}.

In this work, we have shown that VSR's shape-recovery strain $\epsilon(t)$, measured after a step in strain $\epsilon_{0}$ between times $t_{0}$ and $t_{1}$, is consistent with calculations based on the Fractional Zener model. Although a closed form solution for the Fractional Zener model's strain during the shape recovery period could not be obtained, approximate solutions to the model closely resemble the measured behaviors of VSR.

We find that when the compression times are relatively short the fractional order $B$ of the measured shape-recovery process in VSR is similar to but not identical to the fractional order $\beta$ of the measured stress relaxation process in VSR. It is likely, however, that in the limit of infinite compression times $B$ and $\beta$ become identical.

We find that the characteristic time $T$ for the measured shape-recovery process in VSR is always considerably longer than characteristic time $\tau_{1}$ of the measured stress relaxation process in VSR. For short compression times, $T$ increases approximately in proportion to the square-root of the compression time, but it eventually starts to roll over at longer compression times. For the longest compression times, $T$ stops increasing significantly and probably approaches a limiting value $\tau_{2}$, which can be calculated from the Fractional Zener model and the measured stress relaxation modulus. We find a simple relationship between the characteristic time $\tau_{1}$ for stress relaxation under the Fractional Zener model and the characteristic time $\tau_{2}$ for shape-recovery under that same model.

We note that the discussion in this article applies equally well to uncured borosilicone, which is essentially VSR with no permanent shape. With no static modulus, a borosilicone can be modeled by the simpler Fractional Maxwell model\cite{bloomfield2018} rather than the Fractional Zener model, but the latter is equally acceptable when $E_{1}$ is set to zero.

The analysis presented in this article therefore also applies to borosilicones, including common ones (e.g. Silly Putty) and the more complicated borosilicones of Ref. \cite{bloomfield2018}. With $E_{1}=0$, the relationships become simpler, but key ones do not disappear altogether. Most importantly, the shape-recovery process discussed above and in Section C is still present as long as the fractional order $\beta$ of the Fractional Zener/Maxwell model is non-integer.

The memory introduced by non-integer derivatives plays such a fundamental role in the shape-recovery process that shape-recovery exists in fractional-order systems even when there is no permanent shape to recover. While simple borosilicones with $\beta\approx 1$ have little memory and minimal shape-recovery, complicated borosilicones with $\beta < 1$ do. In future work, we will examine shape-recovery in those special borosilicones.

The author acknowledges useful discussions with Rudy McEntire.

\section{A. Solution to ${}_{a}\textbf{D}^{\alpha}_{x} f(x) = b f(x)$}

We can solve the fractional differential equation
\begin{align}
{}_{a}\textbf{D}^{\alpha}_{x} f(x) = b f(x),
\end{align}
where $1 \ge \alpha \ge 0$, for the initial value $f(a)=A$ by expressing $f(x)$ as a power series in $b(x-a)^{\alpha}$:
\begin{align}
f(x) & =\sum_{k=0}^{\infty} c_{k} \left(b (x-a)^{\alpha}\right)^{k}\nonumber\\
& = \sum_{k=0}^{\infty} c_{k} b^{k} (x-a)^{k\alpha}.\label{eq:PSX}
\end{align}
The fractional differential equation is then
\begin{align}
\sum_{k=0}^{\infty} c_{k} b^{k}{}_{a}\textbf{D}^{\alpha}_{x}(x-a)^{k\alpha} = b \sum_{k=0}^{\infty} c_{k} b^{k} (x-a)^{k\alpha}.\label{eq:FDESF}
\end{align}
From \cite{podlubny1998} (2.117),
\begin{align}
{}_{a}\textbf{D}^{\alpha}_{x}(x-a)^{\gamma}=\frac{\Gamma(1+\gamma)}{\Gamma(1+\gamma-\alpha)}(x-a)^{\gamma-\alpha},\label{eq:FDP}
\end{align}
The left side of Eq. \ref{eq:FDESF} is therefore
\begin{align}
&\sum_{k=0}^{\infty} c_{k} b^{k}{}_{a}\textbf{D}^{\alpha}_{x}(x-a)^{k\alpha}\nonumber\\
& = \sum_{k=0}^{\infty} c_{k} b^{k} {}_a\textbf{D}^{\alpha}_{x}(x-a)^{k\alpha}\nonumber\\
& = \sum_{k=1}^{\infty} c_{k} b^{k} \frac{\Gamma(1+k\alpha)}{\Gamma(1+k\alpha-\alpha)} (x-a)^{k\alpha-\alpha}\nonumber\\
& = \sum_{k=1}^{\infty} c_{k} b^{k} \frac{\Gamma(1+k\alpha)}{\Gamma(1+(k-1)\alpha)} (x-a)^{(k-1)\alpha}\nonumber\\
& = \sum_{j=0}^{\infty} c_{j+1} b^{j+1} \frac{\Gamma(1+(j+1)\alpha)}{\Gamma(1+j\alpha)} (x-a)^{j\alpha},\label{eq:DPE}
\end{align}
where the $k=0$ term was set to zero, based on the assumption that ${}_a\textbf{D}^{\alpha}_{x}\textrm{constant}=0$.

The fractional differential equation in series form is thus
\begin{multline}
\sum_{k=0}^{\infty} c_{k+1} b^{k+1} \frac{\Gamma(1+(k+1)\alpha)}{\Gamma(1+k\alpha)} (x-a)^{k\alpha}\\
= b \sum_{k=0}^{\infty} c_{k} b^{k} (x-a)^{k\alpha} 
\end{multline}
Since each power of $(x-a)$ is linearly independent,
\begin{align}
c_{k+1} b^{k+1}\frac{\Gamma(1+(k+1)\alpha)}{\Gamma(1+k\alpha)} =b c_{k} b^{k}\nonumber\\
c_{k+1} b^{k+1} = c_{k} b^{k+1} \frac{\Gamma(1+k\alpha)}{\Gamma(1+(k+1)\alpha)}\nonumber\\
c_{k+1} = c_{k} \frac{\Gamma(1+k\alpha)}{\Gamma(1+(k+1)\alpha)}
\end{align}
Setting c$_{0}= A$, the initial value, and using this recursion relation to obtain $c_{k+1}$ from $c_{k}$ for all $k \ge 1$ gives
\begin{align}
f(x) & = A \sum_{k=0}^{\infty} \frac{\left(b (x-a)^{\alpha}\right)^{k}}{\Gamma(1+k\alpha)}\nonumber\\
& = A E_{\alpha}[b (x-a)^{\alpha}]
\end{align}
where $E_{\alpha}[z]$ is the one-parameter Mittag-Leffler function, member of a class of generalized exponential functions
\begin{align}
E_{\alpha}[z] & = \sum_{k=0}^{\infty}\frac{z^{k}}{\Gamma(\alpha k + 1)}\\
E_{\alpha,\gamma}[z] & = \sum_{k=0}^{\infty}\frac{z^{k}}{\Gamma(\alpha k + \gamma)}.
\end{align}

\section{B. FDE for Fractional Zener model}
Starting with Eqs. \ref{eq:SE1}--\ref{eq:SE5}, we get:
\begin{gather*}
{}_{a}\textbf{D}^{\beta}_{t}\epsilon_{3}(t) = \frac{\sigma_{2}(t)}{F}\\
{}_{a}\textbf{D}^{\beta}_{t}\epsilon(t)={}_{a}\textbf{D}^{\beta}_{t}\epsilon_{2}(t)+{}_{a}\textbf{D}^{\beta}_{t}\epsilon_{3}(t)\\
{}_{a}\textbf{D}^{\beta}_{t}\epsilon(t)={}_{a}\textbf{D}^{\beta}_{t}\epsilon_{2}(t)+\frac{\sigma_{2}(t)}{F}\\
\sigma_{2}(t)=\sigma(t)-E_{1}\epsilon(t)\\
{}_{a}\textbf{D}^{\beta}_{t}\epsilon(t)={}_{a}\textbf{D}^{\beta}_{t}\epsilon_{2}(t)+\frac{\sigma(t)-E_{1}\epsilon(t)}{F}\\
\epsilon_{2}(t)=\frac{\sigma(t)-E_{1}\epsilon(t)}{E_{2}}\\
{}_{a}\textbf{D}^{\beta}_{t}\epsilon(t)={}_{a}\textbf{D}^{\beta}_{t}\left(\frac{\sigma(t)-E_{1}\epsilon(t)}{E_{2}}\right)+\frac{\sigma(t)-E_{1} \epsilon(t)}{F}\\
{}_{a}\textbf{D}^{\beta}_{t}\epsilon(t)=\frac{1}{E_{2}}{}_{a}\textbf{D}^{\beta}_{t}\sigma(t)-\frac{E_{1}}{E_{2}}{}_{a}\textbf{D}^{\beta}_{t}\epsilon(t)
+\frac{1}{F}\sigma(t)-\frac{E_{1}}{F}\epsilon(t)\\
F {}_{a}\textbf{D}^{\beta}_{t}\epsilon(t)=\frac{F}{E_{2}}{}_{a}\textbf{D}^{\beta}_{t}\sigma(t)-\frac{F E_{1}}{E_{2}}{}_{a}\textbf{D}^{\beta}_{t}\epsilon(t)
+\sigma(t)-E_{1}\epsilon(t)\\
\sigma(t)+\frac{F}{E_{2}}{}_{a}\textbf{D}^{\beta}_{t}\sigma(t)=E_{1}\epsilon(t) + F \left(\frac{E_{2}+E_{1}}{E_{2}}\right){}_{a}\textbf{D}^{\beta}_{t}\epsilon(t)\\
\end{gather*}

\section{C. Complete Shape Recovery Solution}
When the Fractional Zener model is unstress and unstrained prior to a sudden step in strain $\epsilon_{0}$ at time $t_{0}$, held at that strain, and then suddenly freed from stress at time $t_{1}$, Eq. \ref{eq:FZMDE} can be used to obtain strain $\epsilon(t)$ as the model recovers its original unstressed, unstained state during the period $t \ge t_{1}$.

Stress $\sigma(t)$ is zero except during the period $t_{0} \le t < t_{1}$, when it is given by Eq. \ref{eq:SRSol}. An expression for $\sigma(t)$ at any time $t$, making use of the Heaviside step function $H(t)$, appears in Eq. \ref{eq:SigmaHeavi}.

Strain $\epsilon(t)$ is zero before $t_{0}$, $\epsilon_{0}$ during the period $t_{0} \le t < t_{1}$, and as yet unknown during the period $t \ge t_{1}$. To obtain $\epsilon(t)$ during that shape-recovery period, we rewrite Eq. \ref{eq:FZMDE} as:

\begin{gather}
{}_{a}\textbf{D}^{\beta}_{t}\epsilon(t)+\frac{E_{1}E_{2}}{F(E_{2}+E_{1})}\epsilon(t)\hspace{4 cm}\nonumber\\
\hspace{1.5 cm} =\frac{E_{2}}{F(E_{2}+E_{1})}\sigma(t)+\frac{1}{E_{2}+E_{1}}{}_{a}\textbf{D}^{\beta}_{t}\sigma(t)\nonumber\\
\end{gather}
Defining
\begin{gather}
\lambda_{2}=-\frac{E_{1}E_{2}}{F(E_{1}+E_{2})}\nonumber\\
A=\frac{E_{2}}{F(E_{2}+E_{1})}\nonumber\\
B=\frac{1}{E_{2}+E_{1}},\nonumber
\end{gather}
this equation becomes
\begin{gather}
{}_{a}\textbf{D}^{\beta}_{t}\epsilon(t)-\lambda_{2}\epsilon(t)
=A\sigma(t)+B{}_{a}\textbf{D}^{\beta}_{t}\sigma(t)\label{eq:FDESREp}
\end{gather}
Equation \ref{eq:FDESREp} is of the form of Eq. \ref{eq:Pod43}, which has solution Eq. \ref{eq:Pod43S}. Using that solution, along with Eq. \ref{eq:SigmaHeavi}, gives
\begin{align}
\epsilon(t) & =\epsilon_{0}\int_{a}^{t}d\tau(t-\tau)^{\beta-1}E_{\beta,\beta}[\lambda_{2}(t-\tau)^{\beta}]\nonumber\\
& \hspace{1.0 cm}\times \Bigg(A\left(E_{1}+E_{2}E_{\beta}\left[-\left(\frac{t-t_{0}}{\tau_{1}}\right)^{\beta}\right]\right)
	W(t)\nonumber\\
& + B{}_{a}\textbf{D}^{\beta}_{t}
	\left(E_{1}+E_{2}E_{\beta}\left[-\left(\frac{t-t_{0}}{\tau_{1}}\right)^{\beta}\right]\right)
	W(t)\Bigg)
\end{align}
where $W(t)=H(t-t_{0})(1-H(t-t_{1}))$.
The integral contains the sum of four terms, which can be write out as the sum of four separate integrals: 
\begin{align}
\epsilon(t) &
=\epsilon_{0}\int_{a}^{t}d\tau(t-\tau)^{\beta-1}E_{\beta,\beta}[\lambda_{2}(t-\tau)^{\beta}]AE_{1}W(\tau)\nonumber\\
& + \epsilon_{0}\int_{a}^{t}d\tau(t-\tau)^{\beta-1}E_{\beta,\beta}[\lambda_{2}(t-\tau)^{\beta}]\nonumber\\
& \hspace{1.0 cm}\times AE_{2}E_{\beta}\left[-\left(\frac{\tau-t_{0}}{\tau_{1}}\right)^{\beta}\right]
W(\tau)\nonumber\\
& + \epsilon_{0}\int_{a}^{t}d\tau(t-\tau)^{\beta-1}E_{\beta,\beta}[\lambda_{2}(t-\tau)^{\beta}]B{}_{a}\textbf{D}^{\beta}_{\tau}E_{1}W(\tau)\nonumber\\
& + \epsilon_{0}\int_{a}^{t}d\tau(t-\tau)^{\beta-1}E_{\beta,\beta}[\lambda_{2}(t-\tau)^{\beta}]\nonumber\\
& \hspace{1.0 cm}\times B{}_{a}\textbf{D}^{\beta}_{\tau}E_{2}E_{\beta}\left[-\left(\frac{\tau-t_{0}}{\tau_{1}}\right)^{\beta}\right]
W(\tau)).
\end{align}
Using the definition of the Riemann-Liouville fractional derivative, Eq. \ref{eq:RLLS}, gives
\begin{align}
\epsilon(t) &
=AE_{1}\epsilon_{0}\int_{t_{0}}^{t_{1}}d\tau(t-\tau)^{\beta-1}E_{\beta,\beta}[\lambda_{2}(t-\tau)^{\beta}]\nonumber\\
& + AE_{2}\epsilon_{0}\int_{t_{0}}^{t_{1}}d\tau(t-\tau)^{\beta-1}E_{\beta,\beta}[\lambda_{2}(t-\tau)^{\beta}]\nonumber\\
& \hspace{1.0 cm}	\times E_{\beta}\left[-\left(\frac{\tau-t_{0}}{\tau_{1}}\right)^{\beta}\right]\nonumber\\
& + \frac{BE_{1}\epsilon_{0}}{\Gamma(1-\beta)}\int_{t_{0}}^{t}d\tau(t-\tau)^{\beta-1}E_{\beta,\beta}[\lambda_{2}(t-\tau)^{\beta}]\nonumber\\
& \hspace{1.0 cm}\times \frac{d}{d\tau}\int_{t_{0}}^{\tau}d\xi(\tau-\xi)^{-\beta}W(\xi)\nonumber\\
& + \frac{BE_{2}\epsilon_{0}}{\Gamma(1-\beta)}\int_{t_{0}}^{t}d\tau(t-\tau)^{\beta-1}E_{\beta,\beta}[\lambda_{2}(t-\tau)^{\beta}]\nonumber\\
& \hspace{0.5 cm}\times \frac{d}{d\tau}\int_{t_{0}}^{\tau}d\xi(\tau-\xi)^{-\beta}E_{\beta}\left[-\left(\frac{\xi-t_{0}}{\tau_{1}}\right)^{\beta}\right]
W(\xi)).
\end{align}
Breaking two of the integrals into pairs of integrals eliminates the $W(\xi)$ functions and gives
\begin{align}
\epsilon(t) & =AE_{1}\epsilon_{0}\int_{t_{0}}^{t_{1}}d\tau(t-\tau)^{\beta-1}E_{\beta,\beta}[\lambda_{2}(t-\tau)^{\beta}]\nonumber\\
& + AE_{2}\epsilon_{0}\int_{t_{0}}^{t_{1}}d\tau(t-\tau)^{\beta-1}E_{\beta,\beta}[\lambda_{2}(t-\tau)^{\beta}]\nonumber\\
& \hspace{1.5 cm} \times E_{\beta}\left[-\left(\frac{\tau-t_{0}}{\tau_{1}}\right)^{\beta}\right]\nonumber\\
& + \frac{BE_{1}\epsilon_{0}}{\Gamma(1-\beta)}\int_{t_{0}}^{t_{1}}d\tau(t-\tau)^{\beta-1}E_{\beta,\beta}[\lambda_{2}(t-\tau)^{\beta}]\nonumber\\
& \hspace{1.5 cm} \times \frac{d}{d\tau}\int_{t_{0}}^{\tau}d\xi(\tau-\xi)^{-\beta}\nonumber\\
& + \frac{BE_{1}\epsilon_{0}}{\Gamma(1-\beta)}\int_{t_{1}}^{t}d\tau(t-\tau)^{\beta-1}E_{\beta,\beta}[\lambda_{2}(t-\tau)^{\beta}]\nonumber\\
& \hspace{1.5 cm} \times \frac{d}{d\tau}\int_{t_{0}}^{t_{1}}d\xi(\tau-\xi)^{-\beta}\nonumber\\
& + \frac{BE_{2}\epsilon_{0}}{\Gamma(1-\beta)}\int_{t_{0}}^{t_{1}}d\tau(t-\tau)^{\beta-1}E_{\beta,\beta}[\lambda_{2}(t-\tau)^{\beta}]\nonumber\\
& \hspace{0.5 cm} \times \frac{d}{d\tau}\int_{t_{0}}^{\tau}d\xi(\tau-\xi)^{-\beta}E_{\beta}\left[-\left(\frac{\xi-t_{0}}{\tau_{1}}\right)^{\beta}\right]\nonumber\\
& + \frac{BE_{2}\epsilon_{0}}{\Gamma(1-\beta)}\int_{t_{1}}^{t}d\tau(t-\tau)^{\beta-1}E_{\beta,\beta}[\lambda_{2}(t-\tau)^{\beta}]\nonumber\\
& \hspace{0.5 cm} \times \frac{d}{d\tau}\int_{t_{0}}^{t_{1}}d\xi(\tau-\xi)^{-\beta}E_{\beta}\left[-\left(\frac{\xi-t_{0}}{\tau_{1}}\right)^{\beta}\right]\label{eq:SRFS}
\end{align}

We are only able to reduce Eq. \ref{eq:SRFS} a little further. A few of the individual integrals have closed-form solutions, but we have not been able to find solutions to the remaining integrals. We leave it to the readers to find a closed form solution to this expression for the strain $\epsilon(t)$ as the Fractional Zener model recovers from a step in strain $\epsilon_{0}$ between $t_{0}$ and $t_{1}$.

\bibliography{shaperecovery}

\end{document}